\documentclass[final,5p,times,twocolumn]{elsarticle}
\usepackage{lineno,hyperref,amsmath,amsthm,amssymb,amsfonts,ragged2e,color,subfig}
\journal{``High Energy Density Physics"}
\begin{document}
\begin{frontmatter}
\title{Electrostatic rogue waves in double pair plasmas}
\author{N. Ahmed$^*$, A. Mannan, N. A. Chowdhury, and A. A. Mamun}
\address{Department of Physics, Jahangirnagar University, Savar, Dhaka-1342, Bangladesh\\
Email: $^*$nahmed93phy@gmail.com}
\begin{abstract}
  A  nonlinear Schr\"{o}dinger equation is derived to investigate the modulational instability (MI)
of ion-acoustic (IA) waves (IAWs) in a double pair plasma system containing adiabatic positive and negative ion fluids along with super-thermal
electrons and positrons. The analytical analysis predicts two types of modes, viz.
fast ($\omega_f$) and slow ($\omega_s$) IA modes. The possible stable and unstable parametric regions for the
IAWs in presence of external perturbation can be observed for both $\omega_f$ and  $\omega_s$.
The number density of the negative ions and positrons play a vital
role in generating the IA rogue waves (IARWs) in the modulationally unstable region. The applications of our present
work  in astrophysical
environments [viz. D-region ($\rm H^+, O_2^-$) and F-region ($\rm H^+, H^-$) of the Earth's
ionosphere] as well as in laboratory plasmas [viz. pair-ion Fullerene ($\rm C^+, C^-$)] are pinpointed.
\end{abstract}
\begin{keyword}
NLSE \sep modulational instability \sep rogue waves.
\end{keyword}
\end{frontmatter}
\section{Introduction}
\label{2:Introduction}
The study of double pair plasma (DPP) has received an enormous attention in plasma physics research community due to their
ubiquitous existence in astrophysical environments, viz., upper regions of Titan's atmosphere \cite{El-Labany2012},
D-regions ($\rm H^+, O_2^-$) and F-regions ($\rm H^+, H^-$) of the Earth's ionosphere \cite{Elwakil2010}  as well as potential diverse application in
laboratory experiments, viz., Fullerene ($\rm C^+, C^-$) \cite{Sabry2008,Oohara2003a,Oohara2003b}, neutral beam sources \cite{Bacal1979}, plasma processing
reactors \cite{Gottscho1986}, and laboratory experiment ($\rm Ar^+, F^-$) \cite{Jacquinot1977,Abdelwahed2016}, etc. Many researchers
have utilized wave dynamics, namely, ion-acoustic (IA) waves (IAWs) for understanding a variety of nonlinear structures, likely,  shock, soliton,
envelope soliton \cite{Chowdhury2018}, and  rogue waves (RWs), in the DPP medium (DPPM) \cite{El-Labany2012,Elwakil2010,Sabry2008}.

The presence of super-thermal electrons/positrons, which move faster than their thermal speed and observed
by Freja satellite \cite{Louran1994}, in DPPM are provided a great effect to generate various kind of nonlinear
phenomena, namely, modulational instability (MI),  envelope soliton \cite{Chowdhury2018}, and
gigantic waves \cite{Chowdhury2017a}, etc. These super-thermal
electrons/positrons are described  appropriately by the generalized Lorengian or $\kappa$-distribution
functions \cite{Vasyliunas1968,Chowdhury2017a}. This spectral index $\kappa$ in such $\kappa$-distribution
measures the strength of plasma particles. It may be noted that the small values of $\kappa$ represent a ``hard"
spectrum with a long tail while for large values of $\kappa$ (specially  $\kappa\rightarrow\infty$)
represent the usual Maxwellian distributions \cite{Vasyliunas1968,Chowdhury2017a}. Ghosh \textit{et al.} \cite{Ghosh2012}
studied IA solitary waves (IASWs) in presence of  $\kappa$-distributed electrons in a three component
unmagnetized plasma medium, and observed that $\kappa$ leads to decrease in the velocity of IASWs.
Hussain \textit{et al.} \cite{Hussain2012} examined that small values of $\kappa$ enhances the nonlinearity
of the plasma medium and the amplitude of the IASWs. Saha \textit{et al.} \cite{Saha2014} studied
IAWs in presence of  $\kappa$-distributed electrons and positrons in electron-positron-ion (e-p-i) plasmas
and observed that the amplitude of the periodic wave decreases with $\kappa$. Chatterjee \textit{et al.} \cite{Chatterjee2011}
investigated IASWs in e-p-i plasma with $\kappa$-distributed electrons and positrons, and
found that the super-thermality of the  electrons and positrons play a significant role on the collision of IASWs.

During the last few decades, the study of MI and associated nonlinear structures (due to the empirical results support the existence of
envelope soliton \cite{Chowdhury2018} or gigantic waves \cite{Abdelwahed2016,Chowdhury2017a}  in fluid dynamics,
optical fiber, and plasma physics) is one of the
eye-catching topic for researchers. Actually, the MI leads to generate a new kind of high energy and very large amplitude structures
known as RWs and this kind of waves are short-lived phenomenon that emerge from nowhere and disappear without a trace.
Recently, a number of authors have studied the MI and RWs in DPPM. Sabry \cite{Sabry2008} studied MI in pair-ion plasma medium (PIPM) in
presence of dust impurities. Elwakil \textit{et al.} \cite{Elwakil2010} reported the propagation of the IAWs
in PIPM, and found that the non-thermality of the electrons decreases stable domain of the IAWs.
Abdelwahed \textit{et al.} \cite{Abdelwahed2016} studied propagation of IAWs in PIPM, and observed
that ratio of ion mass  plays a vital role to manifest the IA RWs (IARWs) in the presence of
$\kappa$-distributed electrons. El-Labany \textit{et al.} \cite{El-Labany2012} investigated IAWs in three components PIPM in presence
of iso-thermal electrons, and found that the negative ion number density enhances the amplitude of the IARWs.
To the best of our knowledge, the  effects of the $\kappa$-distributed electrons and positrons on the MI of the
IAWs and IARWs in a four component (comprising $\kappa$-distributed electrons and positrons, adiabatic positive
and negative ions) DPPM have not been investigated. Therefore, it is a practical interest to
examine the effects of $\kappa$-distributed electrons/positrons on the MI and IARWs in a four component DPPM.

The rest of the paper is organized in the following fashion: The model equations for the
IAWs in a DPPM  with super-thermal electrons and positrons is presented in Sec. \ref{2:Governing equation}.
The stability of the IAWs is shown in Sec. \ref{1:Stability analysis}.
The discussion is provided in Sec. \ref{1:Discussion}.
\section{Model Equations}
\label{2:Governing equation}
We consider a collisionless unmagnetized four component plasma medium having inertialess
$\kappa$-distributed electrons (mass $m_e$; charge $q_e=-e$) and positrons (mass $m_p$; charge $q_p=e$),
inertial adiabatic  negative ions (mass $m_-$; charge $q_-=-eZ_-$) as well as
adiabatic positive ion (mass $m_+$; charge $q_+=eZ_+$).
Here, $Z_-$ ($Z_+$) is the charge state of negative (positive) ion. The overall charge
neutrality condition at equilibrium can be expressed as $Z_+ n_{+0}+n_{p0}=n_{e0}+Z_-n_{-0}$,
where $n_{+0}$, $n_{p0}$, $n_{-0}$, and $n_{e0}$ are the equilibrium number densities of
positive ions, super-thermal positrons, negative ions, and super-thermal electrons, respectively.
Now, the dynamics of the plasma system can be expressed by the dimensionless form of the basic equations as:
\begin{eqnarray}
&&\hspace*{-1.3cm}\frac{\partial n_-}{\partial t} + \frac{\partial}{\partial x} (n_- u_-)=0,
\label{2eq:1}\\
&&\hspace*{-1.3cm}\frac{\partial u_-}{\partial t} + u_-\frac{\partial u_- }{\partial x}+3 \sigma_1 n_- \frac{\partial n_-}{\partial x}=\frac{\partial \phi}{\partial x},
\label{2eq:2}\\
&&\hspace*{-1.3cm}\frac{\partial n_+}{\partial t} + \frac{\partial}{\partial x} (n_+ u_+)=0,
\label{2eq:3}\\
&&\hspace*{-1.3cm}\frac{\partial u_+}{\partial t} + u_+\frac{\partial u_+ }{\partial x}+3 \sigma_2 n_+ \frac{\partial n_+}{\partial x}=-\alpha \frac{\partial \phi}{\partial x},
\label{2eq:4}\\
&&\hspace*{-1.3cm}\frac{\partial^2 \phi}{\partial x^2}=(\mu+\mu_p -1)n_e-\mu_p n_p + n_- -\mu n_+,
\label{2eq:5}\
\end{eqnarray}
The normalizing and associated parameters are represented as: $n_-=N_-/n_{-0}$, $n_+=N_+/n_{+0}$,
$n_e=N_e/n_{e0}$, $n_p=N_p/n_{p0}$, $u_-=U_-/C_-$, $u_+=U_+/C_-$, $x=X/\lambda_{D-}$, $t=T\omega_{p-}$,
$\phi=e\tilde{\Phi}/k_BT_e$, $C_-=\sqrt{Z_- k_BT_e/m_-}$, $\omega_{p-}=\sqrt{4 \pi e^2 Z_-^2 n_{-0}/m_-}$,
$\lambda_{D-}=\sqrt{k_BT_e/4 \pi e^2 Z_- n_{-0}}$, $P_-=P_{-0}(N_-/n_{-0})^\gamma$, $P_{-0}=n_{-0}k_BT_-$, $P_+=P_{+0}(N_+/n_{+0})^\gamma$,
$P_{+0}=n_{+0}k_BT_+$, $\gamma=(N+2)/N$, $\sigma_1=T_-/Z_- T_e$, $\sigma_2=m_- T_+/Z_- m_+ T_e$, $\mu=Z_+n_{+0}/Z_- n_{-0}$, $\mu_p=n_{p0}/Z_- n_{-0}$,
$\alpha=Z_+ m_-/Z_- m_+$; where $n_-$, $n_+$, $n_e$, $n_p$, $u_-$, $u_+$, $x$, $t$, $\tilde{\Phi}$, $C_-$, $\omega_{p-}$,
$\lambda_{D-}$, $T_-$, $T_+$, $T_e$, $T_p$, $P_{-0}$, and $P_{+0}$ is the number densities of negative ion, positive ion, electron, positron, negative ion fluid speed, positive ion fluid speed, space co-ordinate, time co-ordinate, electro-static wave potential, sound speed of negative ion, negative ion plasma frequency, negative ion Debye length, negative ion temperature, positive ion temperature, electrons temperature, positrons temperature, the equilibrium adiabatic pressure of the negative ions, and the equilibrium adiabatic pressure of the positive ions, respectively. Here, $N$ is the number of degrees of freedom and $N=1$ stands for adiabatic one-dimensional case. In addition, it is important to note that we consider for our numerical analysis $m_+ > m_-$, $n_{+0}>n_{-0}$, and $T_e$, $T_p$ $>$ $T_-$, $T_+$. The normalized number density of the $\kappa$-distributed
 electrons is given by \cite{Vasyliunas1968,Chowdhury2017a}
\begin{eqnarray}
&&\hspace*{-1.3cm}n_e =\Big[1-\frac{\phi}{\kappa-3/2}\Big]^{-\kappa+\frac{1}{2}}
\nonumber\\
&&\hspace*{-0.8cm}=1+n_1 \phi +n_2 \phi^2 +n_3 \phi^3 +\ldots,
\label{2eq:6}
\end{eqnarray}
where
\begin{eqnarray}
&&\hspace*{-1.3cm}n_1 =\frac{(\kappa-1/2)}{(\kappa-3/2)},
\nonumber\\
&&\hspace*{-1.3cm}n_2 =\frac{(\kappa-1/2) (\kappa+1/2)}{2 (\kappa-3/2)^2},
\nonumber\\
&&\hspace*{-1.3cm}n_3 =\frac{(\kappa-1/2) (\kappa+1/2)(\kappa+3/2)}{6 (\kappa-3/2)^3}.
\nonumber\
\end{eqnarray}
It is essential to specify that small values of $\kappa$ represent strong super-thermality and
for a physically acceptable distribution, $\kappa>3/2$ is required. However, in the limit
$\kappa\rightarrow\infty$, the difference amongst kappa and Maxwellian distribution is negligible.
The normalized number density of the $\kappa$-distributed positrons is given by \cite{Vasyliunas1968,Chowdhury2017a}
\begin{eqnarray}
&&\hspace*{-1.3cm}n_p =\Big[1+\frac{\delta\phi}{\kappa-3/2}\Big]^{-\kappa+\frac{1}{2}}
\nonumber\\
&&\hspace*{-0.8cm}=1-n_1 \delta\phi +n_2 \delta^2\phi^2 -n_3 \delta^3\phi^3 +\ldots,
\label{2eq:7}
\end{eqnarray}
where $\delta=T_e/T_p$. Now, by substituting \eqref{2eq:6} and \eqref{2eq:7} into \eqref{2eq:5}, and expanding the equation up to third order, we get
\begin{eqnarray}
&&\hspace*{-1.3cm}\frac{\partial^2 \phi}{\partial x^2}+1+\mu n_+=\mu+n_- +\gamma_1 \phi+\gamma_2\phi^2 +\gamma_3 \phi^3 +\ldots,
\label{2eq:8}\
\end{eqnarray}
where
\begin{eqnarray}
&&\hspace*{-1.3cm}\gamma_1= n_1 (\mu+\mu_p-1+\mu_p\delta),
\nonumber\\
&&\hspace*{-1.3cm}\gamma_2= n_2 (\mu+\mu_p-1-\mu_p\delta^2),
\nonumber\\
&&\hspace*{-1.3cm}\gamma_3= n_3 (\mu+\mu_p-1+\mu_p\delta^3).
\nonumber\
\end{eqnarray}

In order to analyze the one-dimensional electrostatic perturbations propagating in our DPPM, we will
derive the nonlinear Schr\"{o}dinger equation (NLSE) by employing the reductive perturbation method (RPM).
The independent variables are stretched  as $\xi=\epsilon (x-v_gt)$ and $\tau=\epsilon^2t$, where
$\epsilon$ is a small expansion parameter and $v_g$ is the group velocity of the IAWs. All dependent
variables can be expressed in a power series of $\epsilon$ as:
\begin{eqnarray}
&&\hspace*{-1.3cm}\Lambda (x.t)=\Lambda_0+\sum_{m=1}^\infty\epsilon ^{(m)}\sum_{l=-\infty}^\infty \Lambda_{l}^{(m)}(\xi,\tau)~\mbox{exp}(il\Upsilon),
\label{2eq:9}\
\end{eqnarray}
where $\Lambda_{l}^{(m)}=[n_{-l}^m$, $u_{-l}^m$, $n_{+l}^m$, $u_{-l}^m$, $\phi_l^m$], $\Lambda_0=[1, 0, 1, 0, 0]^T$, and
$\Upsilon=kx-wt$. Here, the carrier wave number $k$ and frequency $\omega$ are real variables.
We are going to parallel mathematical steps as Chowdhury \textit{et al.} \cite{Chowdhury2018} have done in their work to find successively
the IAWs dispersion relation, group velocity, and NLSE. The  IAWs dispersion relation
\begin{eqnarray}
&&\hspace*{-1.3cm}\omega^2=\frac{A \pm \sqrt{ A^2 - 4 BD}}{2B}.
\label{2eq:10}
\end{eqnarray}
where
$A=k^2(\lambda_1 k^2 +\gamma_1\lambda_1+\lambda_2 k^2+\gamma_1\lambda_2+1+\alpha \mu)$,
$B=(k^2 +\gamma_1)$, $D= k^4(\lambda_1 \lambda_2 k^2 + \gamma_1\lambda_1 \lambda_2+\lambda_1\alpha \mu +\lambda_2)$,
$\lambda_1=3 \sigma_1$, and $\lambda_2= 3 \sigma_2$.
Equation \eqref{2eq:10} describes both the fast (for positive sign) and slow (for negative sign) IA modes denoted by $\omega_f$ and $\omega_s$, respectively.
It is worth mentioning that the condition $A^2 >4 BD$ should be verified to get real and positive values of $\omega$.
The group velocity $v_g$ of IAWs can be expressed as
\begin{eqnarray}
&&\hspace*{-1.3cm}v_g= \frac{S^2 F_1+ M^2(\lambda_1 k^2 +\omega^2 -S)}{2\omega k(M^2 + \alpha \mu S^2)},
\label{2eq:11}\
\end{eqnarray}
where $F_1=(\alpha \mu \omega^2 -2 M^2 +\alpha \mu M + k^2 \lambda_2\alpha \mu)$, $S=k^2 \lambda_1 -\omega^2$, and $M= \omega^2 - k^2 \lambda_2$. Finally, one can obtain the following NLSE:
\begin{eqnarray}
&&\hspace*{-1.3cm}i \frac{\partial \Phi}{\partial \tau} + P \frac{\partial^2 \Phi}{\partial \xi^2}+ Q |\Phi|^2 \Phi=0,
\label{2eq:12}\
\end{eqnarray}
where $\Phi=\phi_1^{(1)}$ for simplicity. The dispersion coefficient $P$ and the nonlinear coefficient $Q$ are, respectively, given by
\begin{eqnarray}
&&\hspace*{-1.3cm}P= \frac{F_2-S^2 M^2}{2\omega k^2(\alpha \mu S^2 +M^2)},
\label{2eq:13}\\
&&\hspace*{-1.3cm}Q=\frac{F_3}{2 \omega k^2 (\alpha \mu S^2 +M^2)},
\label{2eq:14}\
\end{eqnarray}
where
\begin{figure}[!tbp]
  \centering
  \subfloat[]{\includegraphics[width=70mm, height=40mm]{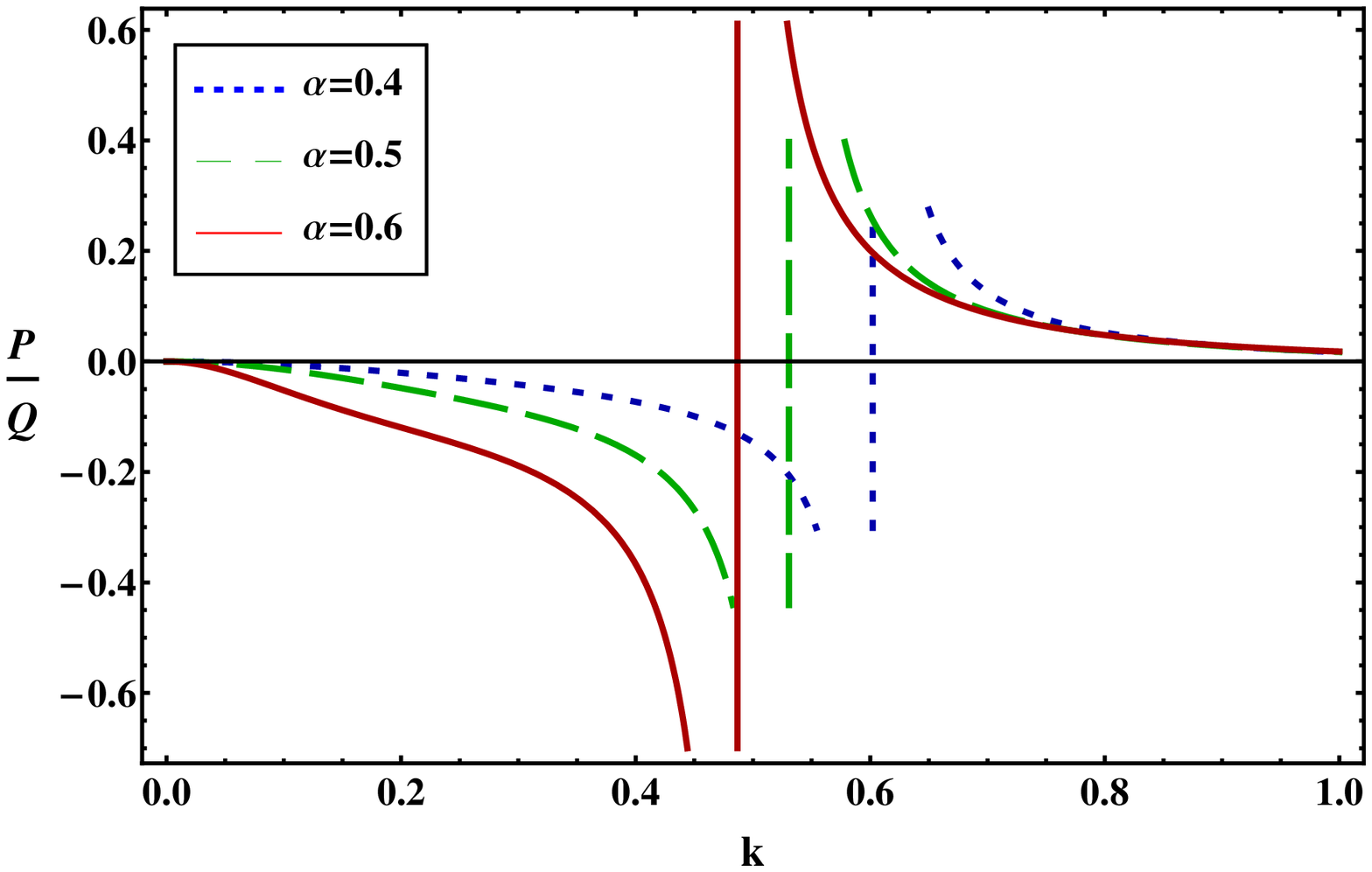}\label{2Fig:F1}}
   \hfill
  \subfloat[]{\includegraphics[width=70mm, height=40mm]{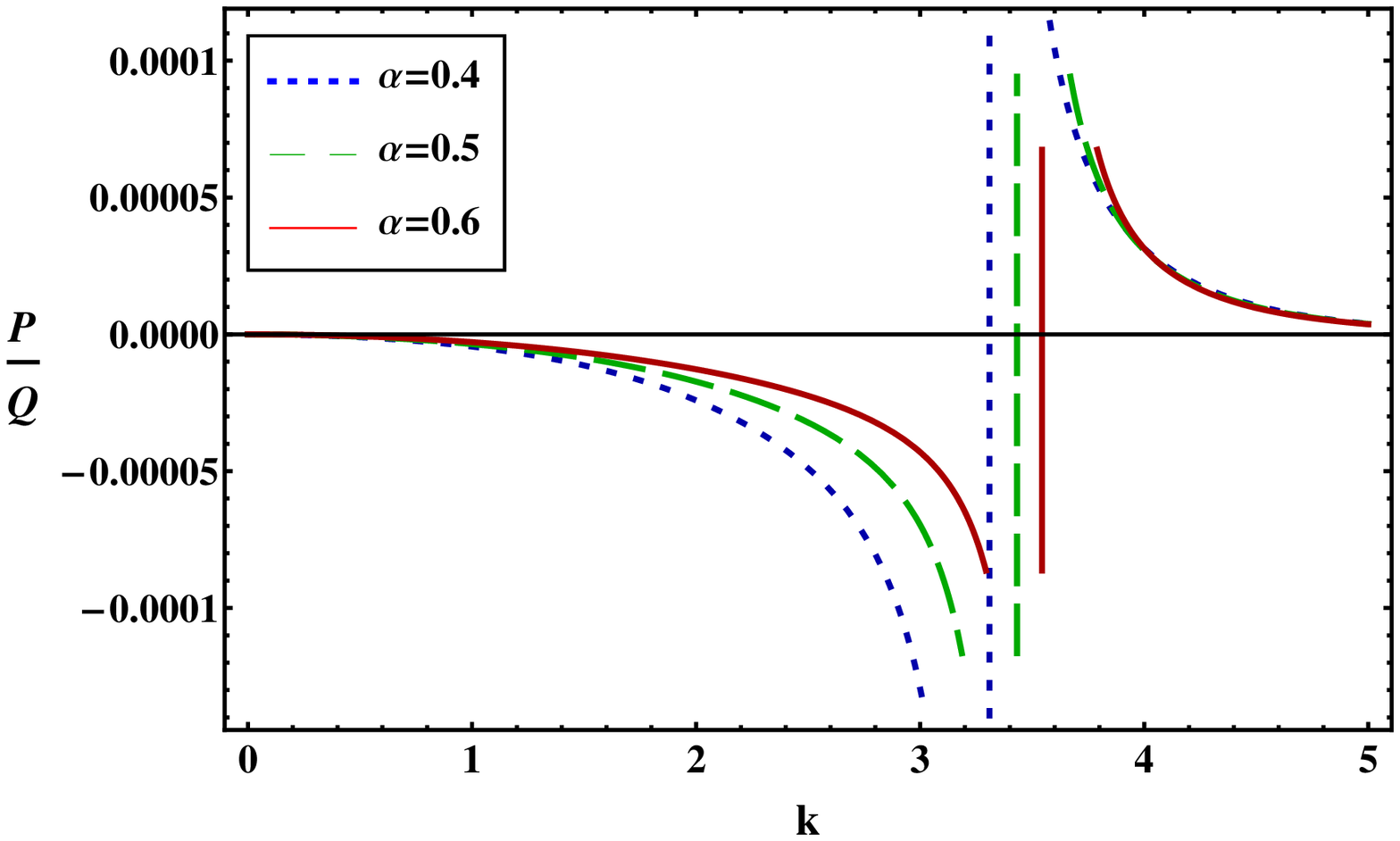}\label{2Fig:F2}}
  \hfill
  \subfloat[]{\includegraphics[width=70mm, height=40mm]{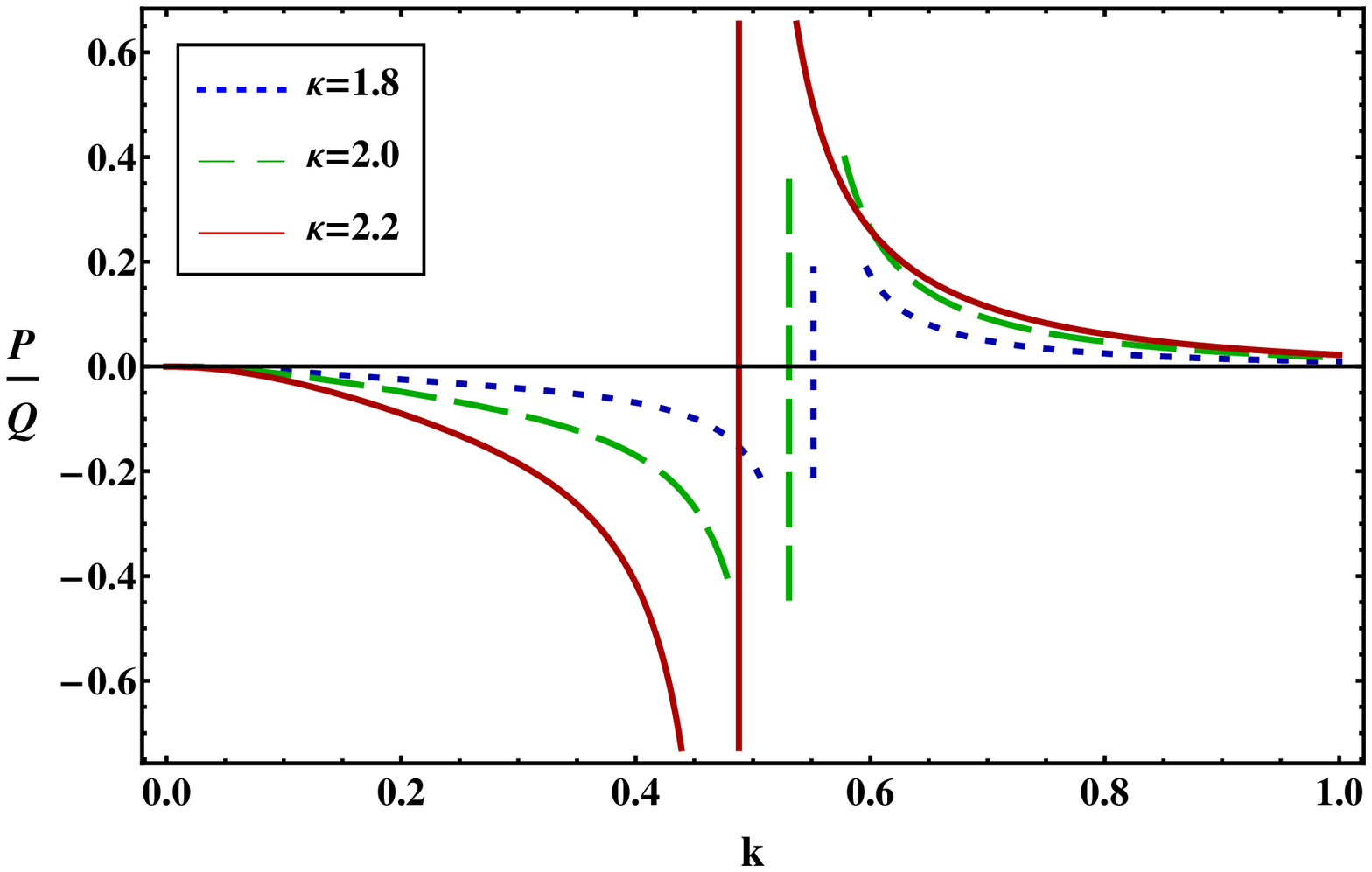}\label{2Fig:F3}}
   \hfill
  \caption{The variation of $P/Q$ with $k$ for different values of (a) $\alpha$ (when $\omega_f$ and $\kappa=2.0$); (b) $\alpha$ (when $\omega_s$ and $\kappa=2.0$); (c) $\kappa$ (when $\omega_f$ and $\alpha=0.5$); along with $\delta=1.3$, $\mu=1.2$, $\mu_p=0.3$, $\sigma_1=0.005$, and $\sigma_2=0.04$.}
\end{figure}
\begin{eqnarray}
&&\hspace*{-1.3cm}F_2=\frac{M^2}{S} [(v_g \omega k-\lambda_1 k^2)( \lambda_1 k^2-2v_g\omega k+\omega^2-S)
\nonumber\\
&&\hspace*{-0.6cm}+(v_g k-\omega)(\omega \lambda_1 k^2 -2v_g k\omega^2 +\omega^3-v_g S k)]
\nonumber\\
&&\hspace*{-0.6cm}-\frac{\alpha \mu S^2}{M}[(v_g \omega k -\lambda_2 k^2)(\lambda_2 k^2 - 2v_g \omega k +M+\omega^2)
\nonumber\\
&&\hspace*{-0.6cm}+(v_g k - \omega)(\omega \lambda_2 k^2 + v_g M k + \omega^3 - 2 v_g k \omega^2)],
\nonumber\\
&&\hspace*{-1.3cm}F_3=3 \gamma_3 M^2 S^2-\alpha \mu k^2 S^2 (\lambda_2 k^2+\omega^2)(C_3 +C_8)
\nonumber\\
&&\hspace*{-0.6cm}+2 \gamma_2 M^2 S^2 (C_5 +C_{10})-2 \alpha \mu \omega S^2 k^3(C_4 +C_9)
\nonumber\\
&&\hspace*{-0.6cm}-k^2 M^2(\lambda_1 k^2+ \omega^2)(C_1 +C_6)-2 \omega M^2 k^3(C_2 +C_7).
\nonumber\
\end{eqnarray}
\begin{eqnarray}
&&\hspace*{-1.3cm}C_1=\frac{k^2 ( 2C_5 S^2 -3 \omega^2 k^2 -\lambda_1 k^4)}{2 S^3},
\nonumber\\
&&\hspace*{-1.3cm}C_2=\Big(\frac{C_1 \omega}{k} -\frac{\omega k^3}{S^2}\Big),
\nonumber\\
&&\hspace*{-1.3cm}C_3=\frac{\alpha k^2 (2 C_5 M^2 + 3 \alpha \omega^2 k^2 +\alpha \lambda_2 k^4 )}{2 M^3},
\nonumber\\
&&\hspace*{-1.3cm}C_4=\frac{\omega (C_3 M^2 -\alpha^2 k^4)}{k M^2},
\nonumber\\
&&\hspace*{-1.3cm}C_5=\frac{ F_4- 2\gamma_2 S^3 M^3}{2 S^3 [M^3 (4k^2+\gamma_1)-\alpha \mu k^2 M^2]+ 2S^2 k^2 M^3},
\nonumber\\
&&\hspace*{-1.3cm} F_4=M^3 k^4 (3 \omega^2 + \lambda_1 k^2)+ \mu S^3 k^4(3 \alpha^2 \omega^2 + \lambda_2 \alpha^2 k^2),
\nonumber\\
&&\hspace*{-1.3cm}C_6=\frac{k^2(\lambda_1 k^2+2 v_g \omega k+\omega^2)-C_{10}S^2}{S^2 (v_g^2 -\lambda_1)},
\nonumber\\
&&\hspace*{-1.3cm}C_7=\frac{C_6 v_g S^2 - 2 \omega k^3}{S^2},
\nonumber\\
&&\hspace*{-1.3cm}C_8=\frac{\alpha^2 k^2(2 v_g \omega k+ \lambda_2 k^2 + \omega^2)+ C_{10} \alpha M^2}{M^2 (v_g^2-\lambda_2)},
\nonumber\\
&&\hspace*{-1.3cm}C_9=\frac{C_8 v_g M^2 - 2 \omega \alpha^2 k^3}{M^2},
\nonumber\\
&&\hspace*{-1.3cm}C_{10}=\frac{2\gamma_2 S^2 M^2 (v_g^2 -\lambda_1)(v_g^2-\lambda_2)+F_5}{S^2 M^2 F_6},
\nonumber\\
&&\hspace*{-1.3cm}F_5= M^2 k^2 (2 v_g k w +\lambda_1 k^2 + \omega^2)(v_g^2-\lambda_2)
\nonumber\\
&&\hspace*{-0.6cm}-\mu S^2 k^2(2v_g \omega k \alpha^2+\lambda_2 \alpha^2 k^2+\alpha^2 \omega^2)(v_g^2-\lambda_1),
\nonumber\\
&&\hspace*{-1.3cm}F_6=[v_g^2 (\alpha \mu +1)-(\alpha \mu \lambda_1+\lambda_2)-\gamma_1(v_g^2-\lambda_1)(v_g^2-\lambda_2)].
\nonumber\
\end{eqnarray}
\section{Stability analysis}
\label{1:Stability analysis}
The sign of the ratio $P/Q$ can recognize the stable/unstable domain for the IAWs in presence of the external perturbations.
The positive (negative) sign of the ratio $P/Q$ defines, respectively, unstable (stable) domain for the IAWs \cite{Sultana2011,Fedele2002,Kourakis2005}.
The intersecting point, in which the stable and unstable domain can be obtain for IAWs, of the $P/Q$ curve
with $k-$axis in $P/Q$ versus $k$ graph is known as critical/threshold wave number  ($k_c$).
We have investigated the stable/unstable domain for the IAWs by depicting $P/Q$ versus $k$ graph for different
values of $\alpha$ and $\kappa$ in Figs. \ref{2Fig:F1}, \ref{2Fig:F2}, and \ref{2Fig:F3} for $\omega_f$, $\omega_s$, and $\omega_f$, respectively.
It can be seen from these figures that (a) the IAWs remain stable for small $k$ ($k<k_c$) and the MI sets in for large values of $k$ ($k>k_c$);
(b) the $k_c$ lies almost in the range of $0.49$ to $0.61$ depending upon the value of $\alpha$ (for fast IA mode) and
the $\alpha$ reduces the stable domain of the IAWs (see Fig. \ref{2Fig:F1}). The variation of $k_c$ with $\alpha$ for slow mode,
exactly an opposite trend is observed with respect to the fast IA mode, can be seen from Fig. \ref{2Fig:F2}.
In this case the $k_c$ bears a value around $3.3$ to $3.6$ (see Fig. \ref{2Fig:F2}).
So, the negative ion mass reduces (enhances) the stable domain of the IAWs corresponds to fast (slow) IA
modes for constant values of positive ion mass ($m_+$), charge state  of the positive ($Z_+$) and negative ($Z_-$) ions (via $\alpha$).
\begin{figure}[!tbp]
  \centering
  \subfloat[]{\includegraphics[width=70mm, height=40mm]{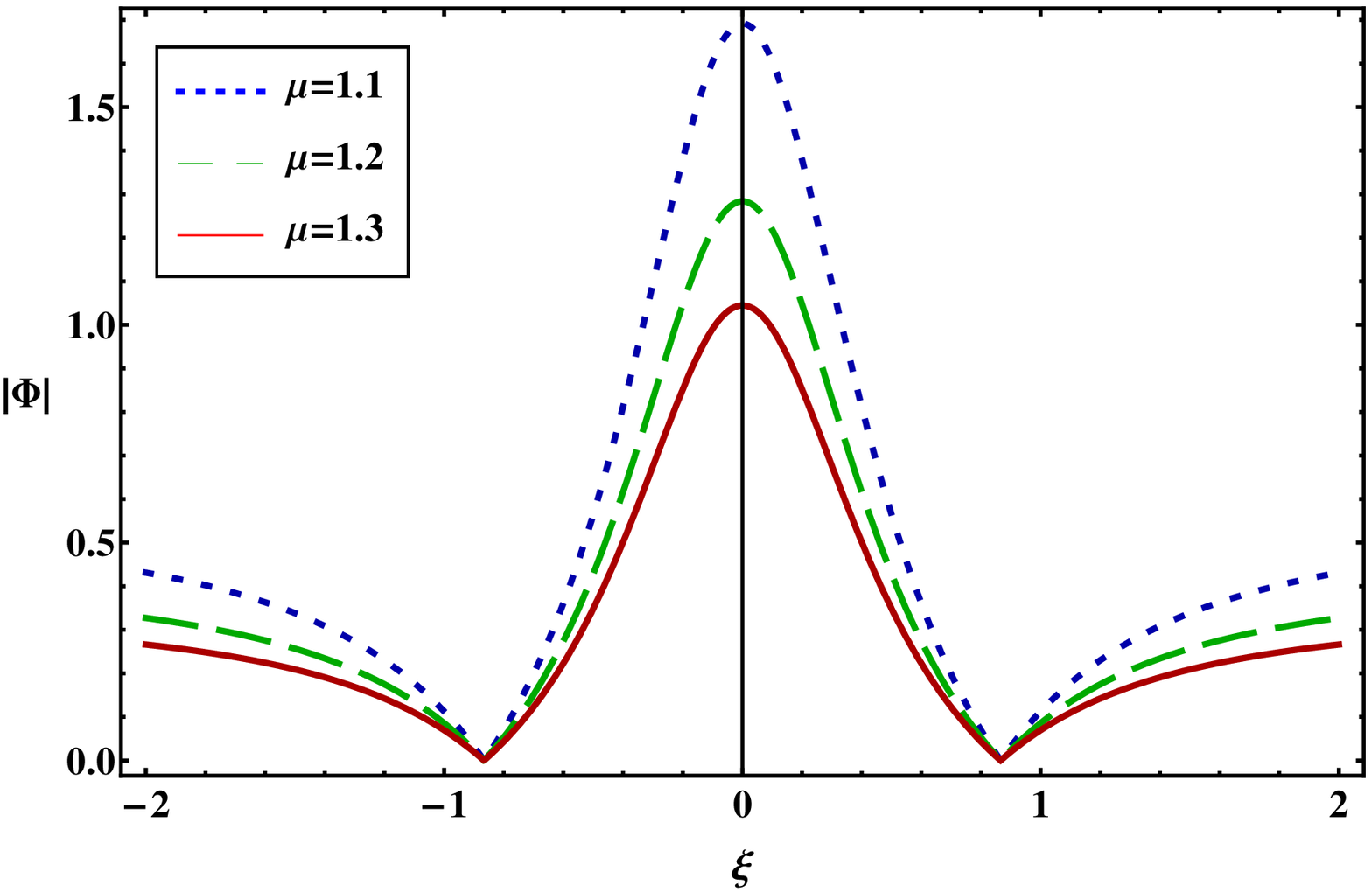}\label{2Fig:F4}}
  \hfill
  \subfloat[]{\includegraphics[width=70mm, height=40mm]{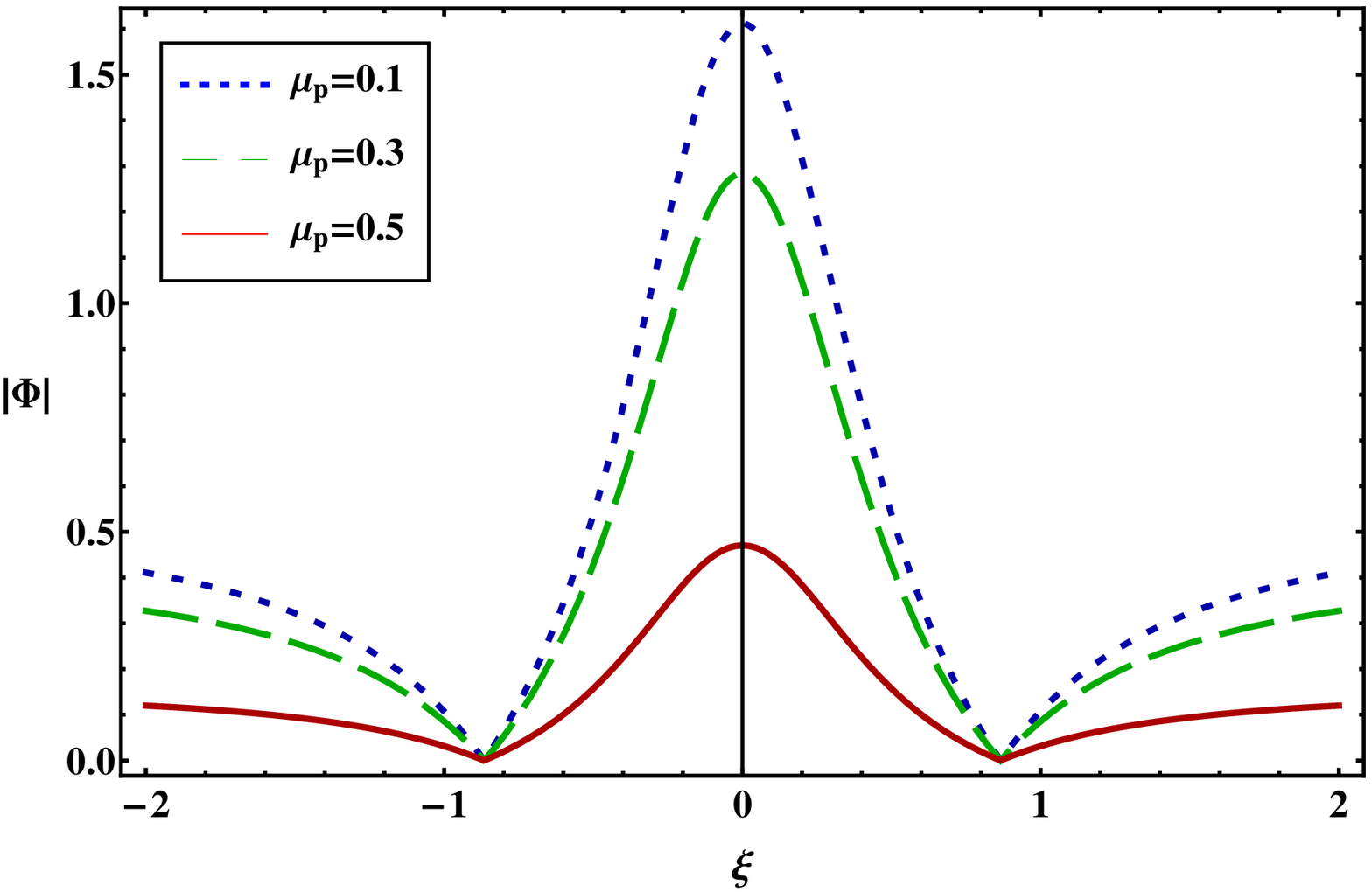}\label{2Fig:F5}}
   \hfill
  \subfloat[]{\includegraphics[width=70mm, height=40mm]{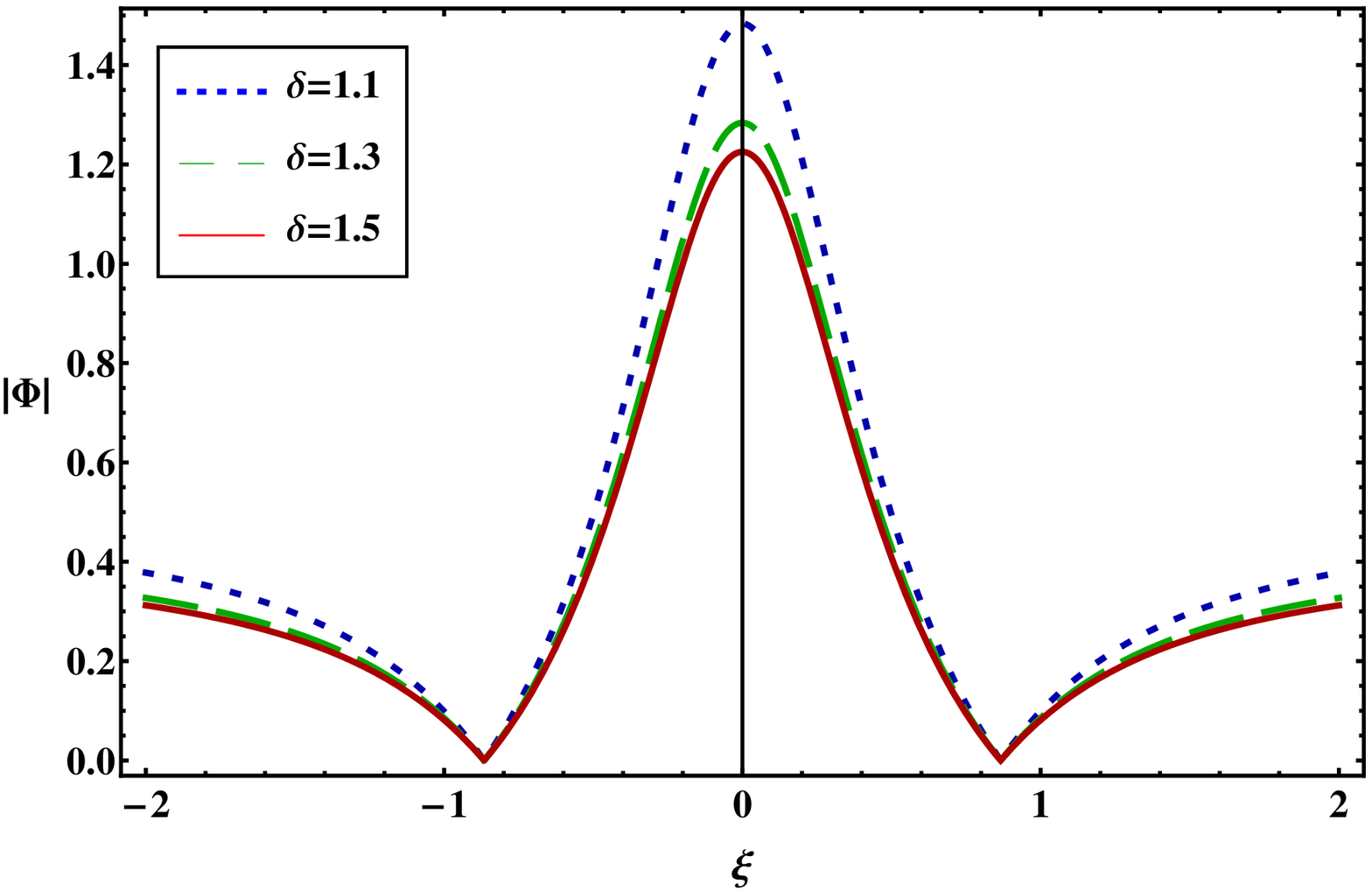}\label{2Fig:F6}}
  \hfill
  \caption{The variation of $|\Phi|$ with $\xi$ for different values of (a) $\mu$ (when $\delta=1.3$ and $\mu_p=0.3$);
   (b) $\mu_p$ (when $\delta=1.3$ and $\mu=1.2$); (c) $\delta$ (when $\mu=1.2$ and $\mu_p=0.3$);
   along with $\alpha=0.5$, $\sigma_1=0.005$, $\sigma_2=0.04$, $\tau=0$, $k=0.7$, and $\omega_f$.}
\end{figure}

The effect of the super-thermality on the stability domain of the IAWs can be observed from
Fig. \ref{2Fig:F3}. This figure shows that by decreasing $\kappa$ leads to an
increase in the  stable domain of IAWs and this result agrees with the result
of Alinejad \textit{et al.} \cite{Alinejad2014} and Gharaee \textit{et al.} \cite{Gharaee2011} works.

In the MI region, a random perturbation in the oscillating ambient background causes
the exponential growth of IAWs amplitude, then rapidly decay without leaving any trace.
This mechanism can be expressed by the first order rational solution or rogue waves, which
developed by the Darboux Transformation Scheme, solution of the NLSE (\ref{2eq:12})
and  can be expressed as \cite{Akhmediev2009,Ankiewiez2009}
\begin{eqnarray}
&&\hspace*{-1.3cm}\Phi(\xi,\tau)=\sqrt{\frac{2P}{Q}}\Big[\frac{4+16 i\tau P}{1+4 \xi^2 + 16\tau^2 P^2}-1\Big] \mbox{exp} (2i\tau P),
\label{2eq:15}
\end{eqnarray}
The solution reveals that a significant amount of IAWs energy is concentrated
into a comparatively small region in DPPM. We are interested to investigate the
effect of various plasma parameters on the features of IARWs [obtained numerically
by using \eqref{2eq:15}]. The effects of the adiabatic positive and negative ion
population, in fact their charge state, on the IARWs can be observed from Fig. \ref{2Fig:F4}
and it is obvious that (a) as we increase the value of $\mu$, the height of the
IARWs decreases; (b) the wave potential increases with
increase in the value of negative ion population ($n_{-0}$), but decreases with
increase of the positive ion population ($n_{+0}$) when $Z_+$ and $Z_-$ remain constant.
Similar fashion is observed from Fig. \ref{2Fig:F5}. In this case, the number density of the
positron (negative ion) minimizes (maximizes) the nonlinearity of the DPPM,
and decreases (increases) the height of the IARWs for constant value of the $Z_-$ (via $\mu_p$).
So, it is clear from both Figs. \ref{2Fig:F4} and \ref{2Fig:F5} that the negative ion population enhances the nonlinearity of the
DPPM in the unstable domain of the IAWs and this is a good agreement with the result of  El-Labany \textit{et al.} \cite{El-Labany2012} work.
Figure \ref{2Fig:F6} discloses the effect of electron and positron temperature on the IARWs (via $\delta$).
The electron temperature reduces the nonlinearity of the DPPM, i.e., the height of the IARWs
decreases, but increases with the increase positron temperature, i.e., the height of the IARWs increases.
\section{Discussion}
\label{1:Discussion}
In summary, a NLSE has been successfully derived to examine and numerically analyzed the MI of IAWs in a DPPM
composed of $\kappa-$distributed electrons and positrons, negatively and positively charged adiabatic ions.
The core results from our present investigation can be summarized as follows:
\begin{enumerate}
\item{The $k_c$, which separates the stable domain from the unstable region, totally depends on the super-thermality of the electrons and positrons, and negative ion mass.}
\item{The negative ion population enhances the nonlinearity of the DPPM in the unstable domain of the IAWs, and increases the height of the IARWs.}
\item{The population of the positron minimizes the nonlinearity of the plasma medium by depicting smaller IARWs.}
\item{The electron (positron) temperature reduces (increases) the height of the IARWs.}
\end{enumerate}
The implications of our results should be useful to understand the nonlinear phenomena (MI and IARWs)
in astrophysical environments, namely, upper regions of Titan's atmosphere \cite{El-Labany2012},
D-region ($\rm H^+, O_2^-$) and F-region ($\rm H^+, H^-$) of the Earth's ionosphere \cite{Elwakil2010}  as well as  in
laboratory experiments, namely, pair-ion Fullerene ($\rm C^+, C^-$) \cite{Sabry2008,Oohara2003a,Oohara2003b}.

\end{document}